\documentclass[12pt]{article}
\usepackage{a4wide}
\usepackage{epsfig}
\def\Journal#1#2#3#4{{#1} {\bf #2}, #3 (#4)}

\def\NPB{{\em Nucl. Phys.} B}

\def\PLB{{\em Phys. Lett.} B}

\def\PRD{{\em Phys. Rev.} D}

\def\ZPC{{\em Z. Phys.} C}

\renewcommand{\slash}[1]{{#1}\!\!\!\!/}

\begin{document}
\thispagestyle{empty}
\begin{flushright}
MZ-TH/98-63\\
hep-ph/9901282\\
January 1999\\
\end{flushright}
\vspace{0.5cm}
\begin{center}
{\Large\bf HEAVY BARYONS --}\\[.5cm]
{\Large\bf STATUS AND OVERVIEW (Theory)}\footnote{Invited 
talk given by J.G. K\"orner at the "8th International Conference
on the Structure of Baryons", Bonn, Sept.22-26, 1998}\\[1.3cm]
\vspace{1.3cm}
{\large S.~Groote$^{1,2}$ and J.G.~K\"orner$^1$}\\[1cm]
$^1$Institut f\"ur Physik, Johannes-Gutenberg-Universit\"at,\\[.2cm]
Staudinger Weg 7, D-55099 Mainz, Germany\\[.5cm]
$^2$Floyd R.~Newman Laboratory of Nuclear Studies,\\[.2cm]
Cornell University, Ithaca, NY 14853, USA
\end{center}
\vspace{1cm}
\begin{abstract}\noindent
We review recent progress in the understanding of the physics of heavy 
baryons. We begin our review by presenting some highlights of recent
experimental findings on charm and bottom baryons and briefly comment on 
their theoretical implications. On the theoretical side we review new
results on the renormalization of HQET, on the Isgur-Wise function for 
$\Lambda_b \rightarrow \Lambda_c$ transitions and on the flavour-conserving
one-pion transitions between heavy baryons.
\end{abstract}

\newpage

\section{Introduction}
Three years ago one of the present authors gave a review talk on heavy
baryons in this series of conferences in Santa Fe, New
Mexico~\cite{kor96a}. Since then there has been considerable experimental 
progress in the study of heavy baryons. Let me just run through
a list of news items from the experimental front: 
\begin{itemize}
\item Altogether 17 charm baryons have been seen to date. This is really
      quite an impressive figure when you compare this 
      to the 13 charm mesons seen so far. This comparison is 
      indicative of the richness of the heavy baryon spectrum as
      compared to the heavy meson spectrum.
\item The new measurement of the polarization of the $\Lambda_b$
      produced on the $Z$-resonance~\cite{opal98} 
      \begin{equation}
      P_{\Lambda_b}= -0.56 {+0.20 \atop -0.13} \pm 0.09
      \end{equation}
      is in better accord with theoretical expectations 
      ($P_{\Lambda_b} \cong -0.6$)~\cite{fape94} than the old (96) ALEPH 
      measurement
      $P_{\Lambda_b}= -.23 { +0.24 \atop -0.20} \pm 0.08$~\cite{aleph96}.
\item An orbitally excited charm-strangeness baryon has been
      seen~\cite{cleo98a}
\item The one-photon transitions $\Xi_c^{\prime} \rightarrow \Xi_c + \gamma$
      have finally been detected~\cite{cleo98b}
\item The life-time of the $\Lambda_b$ is still too small -- and even
      becoming slightly smaller in comparison with the bottom meson
      life times~\cite{wi98}.
\item There are no news on the hyperfine splitting of the 
      Heavy Quark Symmetry doublet $\{ \Sigma_b,\Sigma_b^* \}$. In an
      unpublished '95 paper DELPHI~\cite{delphi95} had reported on a
      disturbingly large hyperfine splitting between the $\Sigma_b^*$
      and the $\Sigma_b$ which has not been confirmed by other experiments.
\item This is unfortunately not a new item. There is presently a large 
      amount of data on heavy baryons  on tapes waiting to be analyzed
      (CLEO, FOCUS, SELEX, SLD, \dots\ ). The community is hopeful that 
      some or all of this data will be analyzed soon.
\end{itemize}
 
A lot of data on heavy baryons can be expected to emerge in the near
future when CLEO III, BaBar, Belle, HERA-B and CDF/D0 (with the new 
injector) begin taking data. Although heavy baryons are not the top priority
of these experiments heavy baryons will certainly be seen if only as 
welcome by-products. Also FOCUS, SELEX and possibly SLD may be back
in action soon. Then there is the European project COMPASS at CERN which
will certainly see charm baryons. In the more distant future there are the
LHC experiments ATLAS, CMS and the dedicated bottom hadron detector LHC-B
as well as the detector BTeV at Fermilab. These next generation
experiments will feature very high bottom quark production rates with
excellent possibilities for the study of bottom baryons and their decays.
  
One may ask what the interest is in studying heavy baryons (charm and bottom)
and the transitions among them? Our favourite answer to this question is
quite simple. A heavy baryon is the ideal place to study the dynamics
of a diquark system in the environment of a heavy quark. In this regard
heavy baryons are far more interesting than heavy mesons where the light
system consists of a single light quark only. Apart from 
some tools designed for the treatment of the heavy degree of freedom (HQET) 
the theoretical analysis of heavy baryons is done with methods well familiar
from the light hadron sector. Among these are lattice simulations,
QCD sum rules, the large $N_C$ limit, chiral perturbation theory, light
cone sum rules, infinite momentum frame techniques, nonrelativistic potential
models, constituent quark models, relativistic quark models, Adler-Weisberger
and Cabibbo-Radicati sum rules, etc.. Lack of space 
prevents us from discussing all of these approaches here.
In the main part of this review we will focus our attention on three topics.
These are the progress
in the renormalization of the HQET Lagrangian, the prospects to determine
the Isgur-Wise function in $\Lambda_b \rightarrow \Lambda_c$
transitions and recent results on one-pion transitions between heavy
baryon states.
 
\section{Progress in the HQET Lagrangian}
The tool to study the physics of heavy baryons and the transitions among 
them is HQET. The HQET Lagrangian is an expansion of the usual QCD 
Lagrangian in terms of inverse powers of the heavy quark mass. In the rest
frame form the HQET Lagrangian reads (see e.g.~\cite{manohar97})
\begin{eqnarray}
{\cal L}_{\rm HQET}=&\psi_Q^\dagger\Big\{iD_0\qquad &\mbox{(static term)}
  \nonumber\\&\qquad
  {\displaystyle +c_k\frac{\vec D^2}{2m_Q}}\qquad
  &\mbox{(kinetic term)}\nonumber\\&\qquad
  {\displaystyle +c_fg\frac{\vec\sigma\cdot\vec B}{2m_Q}}\qquad
  &\mbox{(chromomagnetic term)}\nonumber\\&\qquad
  {\displaystyle +c_dg\frac{[\vec D\cdot\vec E]}{8m_Q^2}}\qquad
  &\mbox{(Darwin term)}\nonumber\\&\qquad
  {\displaystyle +ic_sg\frac{\vec\sigma(\vec D\times\vec E
  -\vec E\times\vec D)}{8m_Q^2}}
  \qquad &\mbox{(spin-orbit term)}\nonumber\\&\qquad
  {\displaystyle +\frac1{m_Q^3}(\mbox{eleven terms})+\,\ldots\,\Big\}\psi_Q}
\end{eqnarray}
where $\vec{E}$ and $\vec{B}$ are the chromoelectric and chromomagnetic 
fields, resp., and where 
$D_{\mu}=\partial_{\mu}-igA_\mu^\alpha T_\alpha=(D_0,-\vec{D})$ is the 
covariant derivative. $\psi_Q$ is the static heavy quark field.

At tree level the coefficients $c_i$ in the HQET Lagrangian are determined
as $c_k=1$, $c_f=1$, $c_d=1$ and $c_s=1$ \cite{manohar97,bkp94}.
The tree level HQET Lagrangian can be obtained from the QCD Lagrangian
through a series of Foldy-Wouthuysen-type transformations
\cite{kt91,bkp94,holstein96}. For example,
after performing the Foldy-Wouthuysen-type transformations up to
${\cal O}(1/m_Q^3)$ one obtains \cite{kt91,bkp94} 
\begin{eqnarray}
{\cal L}_{HQET}^v= && \bar{\psi}_Q \left\{ i\slash{D}_\|
-\frac{1}{2m_Q} \slash{D}_\perp^{\,2}\right.\\
&& \left.
-\frac{i}{4m_Q^2}\left( \frac{1}{2}\slash{D}_\| \slash{D}_\perp^{\,2}
- \slash{D}_\perp \slash{D}_\| \slash{D}_\perp + \frac{1}{2}
\slash{D}_\perp^{\,2} \slash{D}_\| \right)
+ {\cal O}(1/m_Q^3) \right\} \psi_Q\nonumber
\end{eqnarray}
where we have now used the covariant representation of the tree level
HQET Lagrangian. As before $\psi_Q$ is the heavy quark effective 
field. $D_\perp^\mu=D^\mu - v\cdot D v^\mu$ is the transverse component of 
the covariant derivative and
$D_\parallel^\mu=v\cdot D v^\mu$ is its longitudinal 
component where the transverse and longitudinal components are defined
with respect to the arbitrary velocity four-vector $v^\mu=(v_o,\vec{v})$
($v^2=1$).

The HQET Lagrangian possesses a remarkable symmetry, namely 
reparametrization invariance. Reparametrization invariance can be stated
in several equivalent ways. Our favourite way of formulating 
reparametrization invariance is through Lorentz 
invariance. Consider the original QCD Lagrangian (which is 
Lorentz invariant) and expand it to all orders in $1/m_Q$
in terms of two HQET Lagrangians which differ by the velocity
parameter $v$ that specifies them. One has 
\begin{equation} 
\begin{array}{ll}
{\cal L}_{QCD} & =  {\cal L}_{HQET}^v (\mbox{all orders}) \\
              & =  {\cal L}_{HQET}^{v^\prime} (\mbox{all orders})
\end{array}
\end{equation}
It is quite evident that one must have 
\begin{equation}
{\cal L}_{HQET}^v (\mbox{all orders})=
{\cal L}_{HQET}^{v^\prime} (\mbox{all orders})
\end{equation}
On the other hand one can transform ${\cal L}_{HQET}^v$ into
${\cal L}_{HQET}^{v^\prime}$ by the appropiate Lorentz transformation
$v \rightarrow v^\prime$ \cite{blok97}. In this way different coefficients in the HQET 
Lagrangian become related. For example, one has \cite{blok97,chen95}
\begin{eqnarray}
c_k & = & 1 \\
c_s & = & 2 c_f - c_k
\end{eqnarray}
These reparametrization invariance relations are expected to hold to all
orders in $\alpha_s$ of the renormalized HQET Lagrangian. Whether the
conceptually simple derivation of the reparametrization relations
through Lorentz invariance can be upheld to any order of $\alpha_s$ remains
to be seen. The difficulty is that in the derivation \cite{blok97}
it was assumed that the effective fields and operators of the
HQET Lagrangian transform as separate entities under Lorentz 
transformations while they become entangled under renormalization. Needless
to say that the reparametrization relations become very useful 
checks on the correctness of a renormalization calculation. They entail
very powerful identities among loop results even at the one-loop level
which require quite sophisticated means to understand in detail 
\cite{grozin97}.
  
The renormalization of the HQET Lagrangian is achieved through matching
with the corresponding renormalized QCD Lagrangian. The kinetic operator 
does not get renormalized to any loop order. The one-loop renormalization 
of the chromomagnetic operator was first done in \cite{eh90}. The two-loop 
anomalous dimension of the chromomagnetic operator was obtained in
\cite{neubert97} and in \cite{cg97} where the full renormalization
including also finite contributions was carried out. Finite corrections
due to the appearance of two different mass scales were obtained in
\cite{grda98}. The one-loop renormalization of the ${\cal O}(1/m_Q^2)$
operators (Darwin, spin-orbit) was done in \cite{bkp94,ohl96} and in
\cite{blok97,bauer98}
where also the mixing with light quark fields was considered.
The full two-loop renormalization of these operators is in progress
\cite{dggk99}.
Finally I mention first attempts at the one-loop renormalization
of the set of ${\cal O}(1/m_Q^3)$ operators \cite{balz98}.

\section{Isgur-Wise function for
$\Lambda_b \rightarrow \Lambda_c + l^- + \bar{\nu}_l$}
A great deal of experimental and theoretical effort has been expanded on
the determination of the Isgur-Wise function in the exclusive semileptonic
decays of the $B$ meson. Exclusive semileptonic $B$-decays together
with a good understanding of the underlying theory are believed to be one
of the key
experiments in the determination of the KM matrix element $V_{bc}$. It is then
quite natural to ask in what way a $V_{bc}$ determination from exclusive
semileptonic heavy bottom baryon decays could complement the
$V_{bc}$ determination from bottom meson decays. The best candidate
for such a determination certainly is the
$\Lambda_b \rightarrow \Lambda_c$ transition which, in HQET, even
has a simpler structure than the corresponding mesonic transitions.

Let us briefly review the ${\cal O}(1)$ and ${\cal O}(1/m_{Q})$
structure of the $\Lambda_{b}\rightarrow\Lambda_{c}$ form factors as
predicted by HQET (see e.g.~\cite{kkp94}). To leading order in the
heavy mass expansion the $b \rightarrow c$ current matrix element is given by 
\begin{equation}
\langle\Lambda_{c} (v_{2})\mid J_{\mu}^{V-A}\mid\Lambda_{b}(v_{1})\rangle
= F(\omega) \overline{u}_c \gamma_{\mu}(1 - \gamma_{5}) u_b
\end{equation}
where the ${\cal O}(1)$ reduced form factor $F(\omega)$ satisfies the zero
recoil normalization condition $F(\omega = 1) = 1$.
We define the velocity transfer variable $\omega$ by $\omega = v_{1}
\cdot v_{2} $, as usual.

There have been a number of
attempts to calculate the $\omega$-dependence of the Isgur-Wise function
$F(\omega)$. Unfortunately there is a wide spread in the predictions
of the various models for the slope parameter $\rho^2$ characterizing its
fall-off behaviour at zero recoil ($\rho^2$ is defined by the expansion
$F(\omega)=1 - \rho^2(\omega - 1) + \ldots\ $). These range from
$\rho^2 = 1/3$ \cite{ivanov92,ebert96} to $\rho^2$
around 3 \cite{guo93,guo96}. A recent lattice calculation gives a slope of
$\rho^2= 1.2{+0.8\atop-1.1}$ \cite{bowler98}. QCD sum rule determinations
suffer from an
inherent ambiguity resulting from the fact that there is a two-fold 
ambiguity in the choice of
the interpolating fields of the heavy baryon current. For example, in the  
corrected version of \cite{grya92} one has $\rho^2 = 0.85$ for both diagonal
sum rules and $\rho^2 = 0.65$ for the nondiagonal sum rule, both numbers
with a theoretical error of $\approx 0.1$. It is clear that it would be highly
desirable to have some experimental input to clear up the situation.
Unfortunately no data has been published thus far on the baryonic Isgur-Wise
function. The only available result is from a preprint version
of a DELPHI-analysis \cite{delphi97} with the result
$\rho^2 = 1.81 {+0.70 \atop -0.67} \pm 0.32$ which, however, has never 
appeared in print version.

A quick first estimate of the slope of the baryonic Isgur-Wise function
can be obtained by relating it to the mesonic Isgur-Wise function
assuming that the two light quarks in the heavy baryon move independently 
of each other. In this way one
obtains \cite{kkp94,iklr99}
\begin{equation}
F(\omega)= \frac{\omega + 1}{2} \xi^2(\omega)
\end{equation}
where $\xi^2(\omega)$ is the mesonic Isgur-Wise function. The factor
$(\omega +1)/2$ is a purely relativistic effect and guarantees 
the correct threshold behaviour in the crossed
$e^+e^-$-channel \cite{kkp94,iklr99}. For the slope parameter one then obtains 
\begin{equation}
\rho^2_{\rm baryon}=2 \rho^2_{\rm meson} - \frac{1}{2}
\end{equation}
where the term $1/2$ results from the above relativistic effect. Given
that the slope of the mesonic Isgur-Wise function is $\approx 1$ one would
then obtain $\rho^2_{\rm baryon} \approx 1.5$. Incidentally, this value is
quite close to the results of the dipole model in \cite{kkkk97} (1.77), and
the values
of the IMF model \cite{kkkk97} (1.44) and the relativistic three-quark model  
\cite{ikkl97} (1.35) using their favoured sets of model parameters.

At ${\cal O}(1/m_Q)$ all three vector and axial vector form factors
$f_{i}^{V}$ and $f_{i}^{A}$ of the process become populated. They are
defined according to
\begin{eqnarray}
\langle\Lambda_{c} (v_{2})  \mid J_{\mu}^V \mid \Lambda_{b} (v_{1})\rangle
&=& \overline{u}_c(v_2) ( f_1^V \gamma_\mu + f_2^V v_{1\mu} + f_3^V v_{2\mu})
u_b(v_1)
\\
\langle\Lambda_{c} (v_{2})  \mid J_{\mu}^A \mid \Lambda_{b} (v_{1})\rangle
&=& \overline{u}_c(v_2) ( f_1^A \gamma_\mu + f_2^A v_{1\mu} + f_3^A v_{2\mu})
\gamma_5 u_b(v_1)\qquad
\end{eqnarray}
The ${\cal O}(1/m_Q)$ prediction for these form factors read
(see e.g.~\cite{kkp94,kkkk97})
\begin{eqnarray}
  \label{kkp}
  f_1^V(\omega) &=& F(\omega)\; + \;\frac{1}{2}
                            \left[\frac{1}{m_c} + \frac{1}{m_b}\right]\;
                     \left(\eta(\omega) + \bar{\Lambda}F(\omega) \right)
                                                             \nonumber\\
  f_1^A(\omega) &=& F(\omega)\; + \;\frac{1}{2}
                             \left[\frac{1}{m_c} + \frac{1}{m_b}\right]\;
                       \left(\eta(\omega) + \bar{\Lambda}F(\omega)
                           \frac{\omega-1}{\omega+1} \right ) \nonumber\\
  f_2^V(\omega) &=& \quad f_2^A(\omega) =
                     -\frac{1}{m_c}
                      \frac{\bar{\Lambda}F(\omega)}{1+\omega} \nonumber\\
  f_3^V(\omega) &=& - f_3^A(\omega) =
                     -\frac{1}{m_b}
                      \frac{\bar{\Lambda}F(\omega)}{1+\omega}
\end{eqnarray}
where $\eta(\omega)$ satisfies the zero recoil normalization condition
$\eta(\omega=1)=0$. 
Eq.\ (\ref{kkp}) shows that, up to ${\cal O}(1/m_{Q})$, the six form
factors are given in terms of the ${\cal O}(1)$ form factor function
$F(\omega)$,
a new ${\cal O}(1/m_{Q})$ form factor function $\eta(\omega)$ and the 
constant $\bar{\Lambda} \approx M_Q - m_Q \approx 600{\rm\ MeV}$. The 
${\cal O}(1/m_{Q})$ results can be seen to satisfy Luke's theorem which 
reads 
\begin{eqnarray}\label{normv}
f_{1}^{V}(1) + f_{2}^{V}(1) + f_{3}^{V}(1) &=& 1 \nonumber\\
                              f_{1}^{A}(1) &=& 1
\end{eqnarray}

The new unknown ${\cal O}(1/m_{Q})$ form factor function
$\eta(\omega)$ has been found to be negligibly small in two recent theoretical
evaluations using QCD sum rules \cite{dai96} and Infinite Momentum Frame (IMF)
methods \cite{kkkk97}. Some arguments have been presented in \cite{lls98} that
$\eta(\omega)$ is zero at the leading order of the
$1/N_C$-expansion. If one assumes that $\eta(\omega)$ can be entirely
neglected then the ${\cal O}(1/m_{Q})$ behaviour of exclusive
semileptonic $\Lambda_b$ decays is solely determined by the leading order
${\cal O}(1)$ Isgur-Wise function $F(\omega)$ ($\bar{\Lambda}$ can be
determined from elsewhere).
This observation opens the way to a meaningful comparison of 
experimental data analyzed at ${\cal O}(1/m_{Q})$ with theoretical
evaluations done at ${\cal O}(1)$. 

\section {One-pion transitions between heavy baryons}
In this section we will be concerned with flavour-conserving one-pion
transitions between heavy baryons (in contradistinction to flavour-changing
transitions as e.g.\ in $\Lambda_b \rightarrow \Lambda_c + \pi$). We will
discuss ground-state $S$-wave heavy baryons as well as excited $P$-wave
heavy baryons and the one-pion transitions between them. There are two types
of $P$-wave states depending on whether the
two light quarks are in relative $P$-wave or whether the two light
quarks as a whole are in a $P$-wave state relative to the heavy quark. The
latter we call $K$-excitations while we call the former $k$-excitations.
The $K$-excitations lie $\approx 150{\rm\ MeV}$ below the $k$-excitations
according to a potential model calculation using harmonic oscillator
forces~\cite{cik79}. The two experimentally observed excited $\Lambda_c$ 
states $\Lambda_c(2593)$ and
$\Lambda_c(2625)$ are very likely the $J=1/2$ and $J=3/2$ members 
of the lowest lying Heavy Quark Symmetry doublet $\{ \Lambda_{cK1} \}$.
They have been discovered through their pion transitions. In due course
other $P$-wave heavy baryon states and their pion transitions (such as
the evidence for the orbitally excited charm-strangeness baryon
\cite{cleo98a}) will be discovered. It is the purpose of this section to 
describe some of the progress which has been made in the description of 
one-pion transitions between heavy baryons. The languge to be used in this 
description will be the very compact language of the $3nj$-formalism.
 
At the particle level the transition $ J_1 \rightarrow J_2 + \pi(l)$ is
described
in terms of the reduced matrix elements 
\begin{equation}
\langle J_2 \mid\mid {\cal O}^{l}
  \mid \mid J_1 \rangle
\end{equation}
where ${\cal O}^{l}$ is the total transition operator between the
initial state with spin $J_1$ and
the final state
$J_2 + \pi(l)$ with a pion in the orbital state $l$.
To leading order in the heavy quark mass expansion these transitions can be 
viewed as a pion transition between the light diquarks
$ j_1 \rightarrow j_2 + \pi(l)$ in the presence of a noninteracting static
heavy quark. In this limit the transition is thus governed by the reduced
matrix elements
\begin{equation}
\langle j_2 \mid\mid {\cal O}^{l} 
  \mid \mid j_1 \rangle
\end{equation}
Since the number of reduced matrix elements (or coupling constants) at
the diquark level is less than that at the particle level one has achieved
a reduction in the number of independent coupling
constants that describe the one-pion transitions.

The number of independent coupling constants can be further reduced if
one invokes in addition the constituent
quark model for the light-side transitions together with the assumption
that the one-pion transition is a one-body operator (justified in the
$1/N_C$ approach \cite{yan98}). In the constituent quark
model the light-side transitions are given in terms of the product of
reduced matrix elements 
\begin{equation}
\langle s_{q_2} \mid\mid {\cal O}^{\sigma} 
\mid \mid s_{q_1} \rangle \;
\langle L_2 \mid\mid {\cal O}^{L} 
  \mid \mid L_1 \rangle
\end{equation}
where $s_{q_1}$ and $s_{q_2}$ are the active light quarks in the one-pion
transition and ${\cal O}^{\sigma}$ is the spin-$1$ one-body operator that
induces the transition between the two. $L_1$ and $L_2$ are the orbital 
angular momenta of the light quark system and ${\cal O}^{L}$ is the orbital
angular momentum operator that induces the orbital transition.

Technically the reduction from the particle level to the diquark level
and then further to the constituent quark level involves a recoupling analysis
of the various angular momenta involved in the transition. One is
therefore naturally led to the use of $6j$- and $9j$-symbols.

The first stage of the reduction from the particle level to the diquark
level involves a recoupling of the six angular momenta $j_1$ (initial
light diquark spin), $j_2$ (final light diquark spin), $J_1$ (initial
heavy baryon spin), $J_2$ (final heavy baryon spin), $l$ (orbital angular
momentum of pion) and the heavy quark spin $s_Q=1/2$. The number of
angular momenta already suggests that the desired reduction can be achieved
with the help of the $6j$-symbol. In fact one has the two coupling schemes
\begin{eqnarray}
\mbox {\bf{I.}}& \qquad \;\; l+j_2=j_1  \;\;,\;\;   j_1+1/2=J_1 \\
\mbox{\bf{II.}}& \qquad j_2+1/2=J_2 \;\;,\;\;  J_2+l=J_1
\end{eqnarray}
The two coupling schemes are related via recoupling coefficients which
in this case are given by the $6j$-symbol
\begin{eqnarray}
 \left\{\begin{array}{ccc}
     l & j_2 & j_1\\
     1/2 & J_1 & J_2  
  \end{array} \right \}
\end{eqnarray}

\begin{figure}[ht]\begin{center}
\epsfig{figure=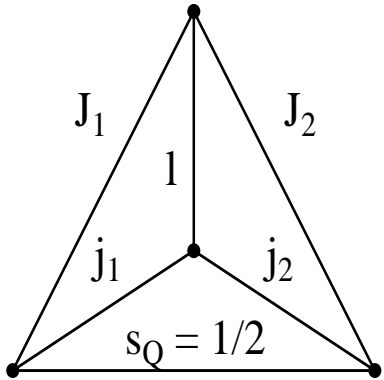, height=5truecm, width=5truecm}
\caption{\label{fig1}Recoupling diagram representing a $6j$-symbol in
the recoupling of six angular momenta in the HQS limit.}
\end{center}\end{figure}

A pictorial representation of the $6j$-symbol is shown in Fig.~\ref{fig1}. 
Each of the six links in Fig.~\ref{fig1} represent angular momenta while 
the four nodes represent the coupling of angular momenta.
  
Using standard orthogonality relations involving C.G. coefficients and the
$6j$-symbol one can relate the reduced matrix elements at the particle and 
diquark level. One has
\begin{eqnarray}
\langle J_2 \mid\mid {\cal O}^{l}\mid \mid J_1 \rangle
  &=&(-1)^{J_1+1/2+l-j_1}\sqrt{(2J_1+1)(2J_2+1)}\nonumber\\&&\times
 \left\{\begin{array}{ccc}
     l & j_2 & j_1\\
     1/2 & J_1 & J_2
  \end{array} \right \}
  \langle j_2 \mid\mid {\cal O}^{l}
    \mid \mid j_1 \rangle
\end{eqnarray}
It is evident that one has thereby achieved a reduction in the coupling
constant complexity. 

In the second stage one resolves the diquark transitions into constituent
quark transitions. At this stage one has to recouple altogether 
twelve angular momenta. The first six are $s_{q_1}$=1/2 and $s_{q_2}$=1/2
(initial and final active light quarks), $s_{q_s}$=1/2 (passive spectator
quark), $S_1$ and $S_2$ (initial and final sum of spins in the light diquark)
and $\sigma$=1 (angular momentum of one-pion transition operator). 
In addition one has the orbital angular momenta $L_1$, $L_2$ and $L$ from 
the orbital transition operator ${\cal O}^L$, and the angular momenta $j_1$,
$j_2$ and $l$, as before.
At first sight one would presume that the one-pion transitions 
are now described in terms of a $12j$-symbol. However, in the constituent quark
model one neglects spin-orbit coupling. The spin and orbital spaces 
decouple and factorise in the transition. This implies that the one-pion
transitions are determined by a product of a $6j$- and $9j$-symbol acting
separately in spin space and orbital space, respectively.

In spin space one has the two coupling schemes
\begin{eqnarray} 
\mbox {\bf{I.}} & \qquad s_{q_s}+s_{q_2}=S_2 \;\; , \;\; S_2+ \sigma= S_1 \\
\mbox{\bf{II.}} & \qquad s_{q_2}+ \sigma= s_{q_1} \;\; , \;\;
  s_{q_s}+s_{q_1}=S_1 
\end{eqnarray}
 with the recoupling coefficient ($6j$-symbol) 
 \begin{eqnarray}
 \left\{\begin{array}{ccc}
     s_{q_s} & s_{q_2} & S_2\\
      \sigma & S_1 & s_{q_1}
  \end{array} \right \}
\end{eqnarray}
In orbital angular momentum space one has the two coupling schemes
\begin{eqnarray}
\mbox{\bf{I.}} & \qquad
   L+ \sigma= l \;\;,\;\;
    L_2+S_2=j_2  \;\;,\;\; l+j_2=
    j_1 \\
\mbox{\bf{II.}} & \qquad
   L +L_2= L _1 
   \;\;,\;\;   \sigma+ S_2=S_1  \;\;,\;\; 
    L_1+S_1=j_1 \; , 
\end{eqnarray}
and thus the recoupling coefficient ($9j$-symbol) 
\begin{eqnarray}
  \left\{
  \begin{array}{rrr}
     L & \sigma & l \\
     L_2 & S_2 & j_2 \\
      L_1 & S_1 & j_1 
  \end{array} 
  \right\}
\end{eqnarray}

\begin{figure}[ht]\begin{center}
\epsfig{figure=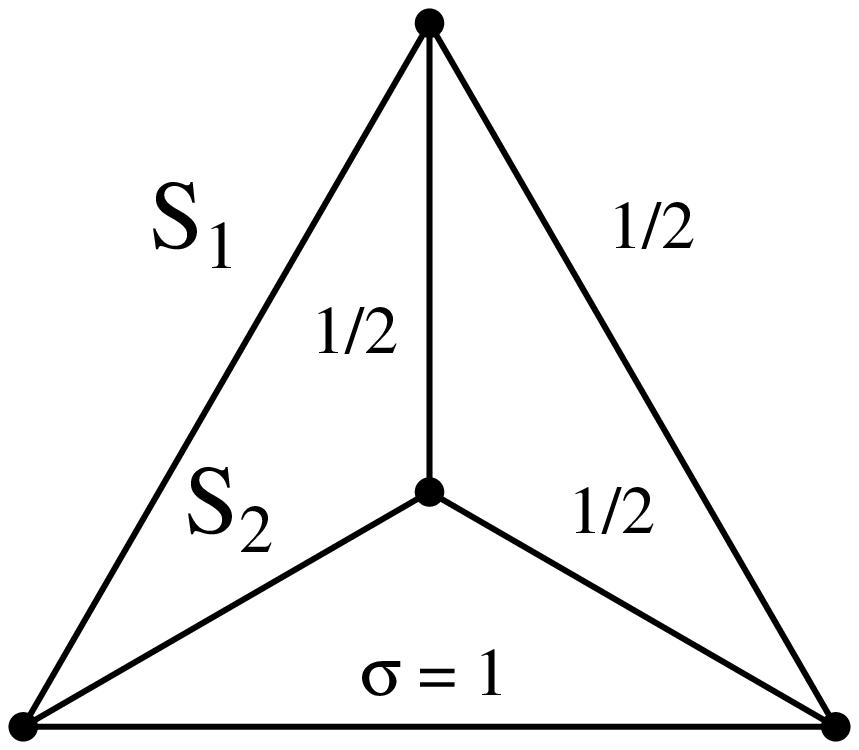, height=5truecm, width=5truecm}\hspace{1truecm}
\epsfig{figure=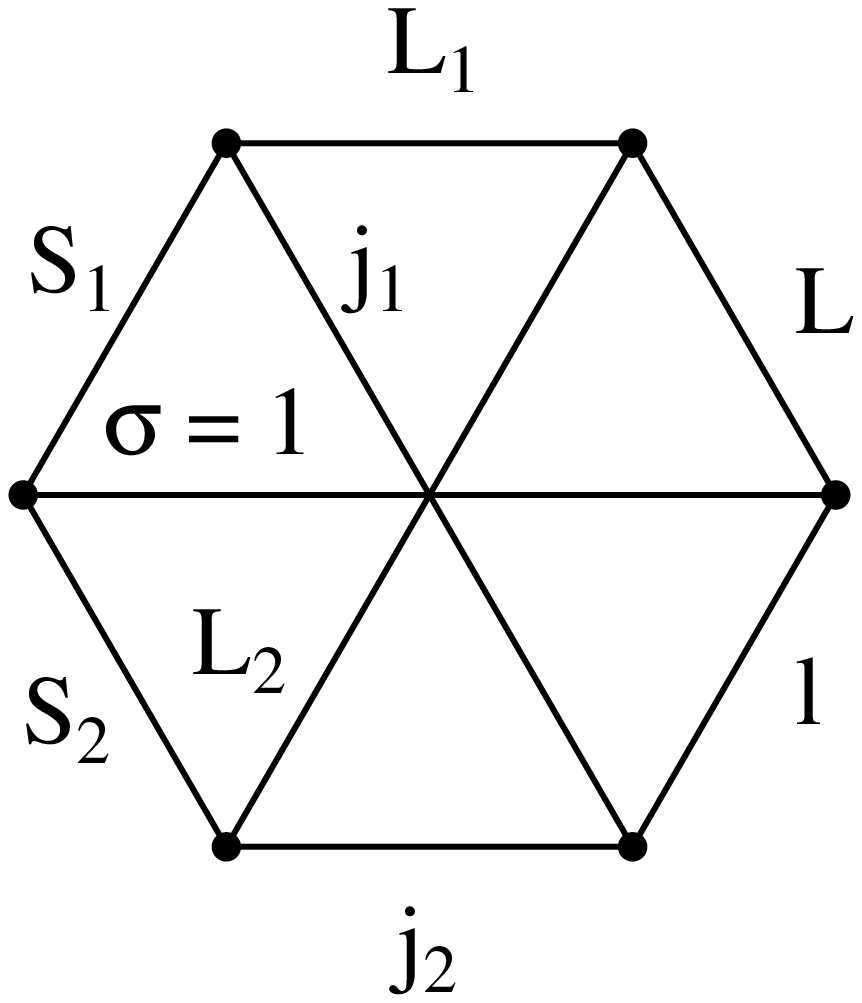, height=6truecm, width=5truecm}\vspace{12pt}
\hbox{\Large\bf(a)\kern6truecm(b)}
\caption{\label{fig2}Recoupling diagrams representing $6j$- and $9j$-symbols
in the constituent quark model approach. a) $6j$-symbol acting in spin-space
b) $9j$-symbol acting in orbital space.}
\end{center}\end{figure}

The relevant recoupling diagrams representing the recoupling of the
respective two coupling schemes are depicted in Fig.~\ref{fig2}. 
Fig.~\ref{fig2}a is a pictorial representation of the $6j$-symbol acting 
in spin space with four nodes (couplings) and six links (angular momenta) 
while Fig.~\ref{fig2}b represents the $9j$-symbol acting in orbital space 
with six nodes and nine links.

The diquark reduced matrix element can thus be expressed in terms of the
product of spin and orbital reduced matrix elements of the constituent
quark model. After a little bit of algebra using identities involving
C.G. coefficients, $6j$- and $9j$-symbols one obtains 
\begin{eqnarray}
 \langle j_2 \mid\mid {\cal O}^{l} \mid \mid j_1 \rangle
  &=&(-1)^{S_{q_s}+S_{q_1}+l+S_1+j_1-j_2}\nonumber\\&&\times
  \sqrt{(2l+1)(2j_1+1)(2j_2+1)(2S_1+1)(2S_2+1)}\label{momenta}\\&&\times 
 \kern-48pt\times\left\{\begin{array}{rrr}
     s_{q_s} & s_{q_1} & S_1\\
      \sigma & S_2 & s_{q_2}  
  \end{array} \right \} 
  \left\{
  \begin{array}{rrr}
     \sigma & L & l \\
     S_1 & L_1 & j_1 \\
      S_2 & L_2 & j_2 
  \end{array} 
  \right\}
  \langle S_{q_2} \mid\mid {\cal O}^{\sigma} \mid \mid S_{q_1} \rangle 
  \langle L_2 \mid\mid {\cal O}^{L} \mid \mid L_1 \rangle \;.\nonumber 
 \end{eqnarray}

This is our master formula giving the predictions of the constituent quark
model for the one-pion transitions between two orbitally excited heavy
baryon states for any general transition $L_1 \rightarrow L_2$ \cite{hkt99}. 
For the two cases discussed here, namely $L_1=L_2=0$ and $L_1=1,\;L_2=0$ 
the structure of the master formula considerably simplifies since the
$9j$-symbol reduces to a Kronecker-$\delta$ in the first case and to a
$6j$-symbol in the second case.

\begin{table}[ht]
\begin{center}
\begin{tabular}{|l|c|c|c|l|}\hline
&\multicolumn{4}{c|}{Number of couplings}\\
&particle level&HQS&CQM&PCAC\\\hline
$S$-wave to $S$-wave&$8$&$2$&$1$&predicted\\
$P$-wave to $S$-wave&$19$&$7$&$2$&\hbox{\qquad}--\\\hline
\end{tabular}
\caption{\label{tab1}Enumeration of the number of independent couplings (or
reduced matrix elements) in one-pion transitions between heavy baryons using
various model assumptions. The $P$-wave states refer either to $K$- or to
$k$-excitations.}
\end{center}
\end{table}
In Table~1 we enumerate the number of independent couplings 
using various model assumptions starting from the particle level. We then 
count the number of independent couplings using
Heavy Quark Symmetry (HQS) and further using the constituent quark
model (CQM). It is gratifying to see how each additional symmetry reduces
the number of independent couplings. In the case of elastic transitions one
even obtains an absolute prediction in the CQM approach since the orbital
overlap becomes normalized in the elastic case and the coupling of the
pion to the constituent quarks is known from PCAC~\cite{yan92}. On the other
hand, in order to obtain
absolute predictions for the $P$-wave to $S$-wave transitions one needs
to bring in further dynamics. First dynamical model calculations in 
this direction
have been done using a light-front quark model~\cite{tawfiq98} and a
relativistic
three-quark model \cite{ivanov99} with some promising results. There is
certainly need
for more data to compare the model predictions with. 

All what was done in this section could have been done using chirally
invariant couplings and
explicit quark model spin wave functions \cite{yan97} or covariant quark model
wave functions \cite{hkt99}. We chose to present our results in terms of the
very elegant $3nj$-symbol approach partly for the reason that parts of
the audience at the Bonn meeting might find it amusing that their 
upbringing in nuclear and/or atomic physics, where
$3nj$-symbols are heavily used, now would allow them to quickly grasp
the physics
of one-pion transitions, and, for that matter, one-photon transitions 
\cite{kkp94,photon99,gkt99} between heavy baryons.

\section{Concluding remarks}
We provided a brief review of a few selected topics in heavy baryon
physics. Lack of space prevented us from covering more details.
We are looking forward to more data on heavy baryons which will hopefully be 
forthcoming soon. These data will certainly stimulate further theoretical
progress.\\[12pt]{\bf Acknowledgments:}
We acknowledge partial support from the BMBF (Germany) under contract 
No.~06MZ865. S.G. gratefully acknowledges a grant given by the Max Kade 
Foundation.

\end{document}